\documentclass[12pt]{article}
\input epsf.tex
\usepackage{amssymb}

\newcommand{\bmat}{\left(\begin{array}}
\newcommand{\emat}{\end{array}\right)}

\def\diagram#1{{\normallineskip=8pt\normalbaselineskip=0pt \matrix{#1}}}

\def\harr#1#2{\smash{\mathop{\hbox to .3in{\rightarrowfill}}
 \limits^{\scriptstyle#1}_{\scriptstyle#2}}}

\def\yzero{\smash{\hbox{$y\kern-4pt\raise1pt\hbox{${}^\circ$}$}}}

\def\s2{\frac{1}{\sqrt2}}

\def\beq{\begin{equation}}
\def\eeq{\end{equation}}
\def\beqa{\begin{eqnarray}}
\def\eeqa{\end{eqnarray}}

\def\Dsl{\,\raise.15ex\hbox{/}\mkern-13.5mu D} 

\def\IR{\mathbb{R}}
\def\IC\mathbb{C}
\def\IZ{\mathbb{Z}}
\def\ItZ{\widetilde{\mathbb{Z}}}
\def\IZdos{{\bf Z}_2}

\def\IS{\mathbb{S}}
\def\S{{\bf S}}

\def\IRP{\mathbb{R}{\rm P}}

\def\RROp{$\widetilde{Op}^{\pm}$}

\def\RROtwo{$\widetilde{O2}^{\pm}$}

\def\RPOp{$\widetilde{Op}^{+}\:$}

\def\RPOtwo{$\widetilde{O2}^{+}\:$}

\def\RNOp{$\widetilde{Op}^{-}\:$}

\def\RNOtwo{$\widetilde{O2}^{-}\:$}

\topmargin
-.5cm
\textwidth
15.5cm
\textheight
23cm
\oddsidemargin
0.7cm
\evensidemargin
1.2cm

\begin{document}

\makeatletter
\@addtoreset{equation}{section}
\makeatother
\renewcommand{\theequation}{\thesection.\arabic{equation}}
\pagestyle{empty}
\rightline{CTP-MIT/3356}

\rightline{\tt hep-th/0303270}
\vspace{.5cm}
\begin{center}
\Large{\bf The Exchange of Orientifold Two-Planes in M-theory}\\

\large
Oscar Loaiza-Brito\footnote{e-mail address:{\tt oloaiza@lns.mit.edu}} \\[2mm]

{\em Center for Theoretical Physics}\\
{\em Massachusetts Institute of Technology}\\ 
{\em Cambridge, Massachusetts 02139, U.S.A.}\\[4mm]

\vspace*{1cm}
\small{\bf Abstract} \\[7mm]
\end{center}

\begin{center} 
\begin{minipage}[h]{14.0cm} {We propose an M-theory lift picture of the exchange among type IIA orientifold two-planes. This consists 
in wrapping a $M5$-brane on a three-cycle in the transverse space of the M-theory orientifold plane $OM2$. 
A flux quantization condition for the
three-form self-dual field strength,  on the worldvolume of the $M5$-brane is computed. This condition establishes the value which explains
the relative charge between two different $OM2$-planes. Also, we find that the exchange of the four  types of 
orientifold two-planes in string theory, has a common picture in M-theory. Moreover, we find that the assignment of the extra charge is 
fixed by cohomology 
and by the flux quantization of the field strength $G_4$ in M-theory. We conclude that
cohomology is sufficient to describe some orientifold properties in M-theory, that at string theory level, only K-theory is able to explain.
}

\end{minipage} 
\end{center}

\bigskip

\bigskip
  

\leftline{November 2003}

\newpage
\setcounter{page}{1}
\pagestyle{plain}
\renewcommand{\thefootnote}{\arabic{footnote}}
\setcounter{footnote}{0}

\section{Introduction}
The study of orientifold planes in string theory is very important basically by two reasons: first, because they offer a framework where
it is possible to construct supersymmetric gauge theories with orthogonal and symplectic groups. Brane setups in the presence
of orientifold planes have allowed us to understand some symmetries and dualities present in string theory. Second, the presence of
orientifold planes implies, as in the case of Type I superstring theory, the existence of non-supersymmetric states known as non-BPS states.
They offer a background where it is possible to construct realistic models where the supersymmetry is broken. 

However, there are also topological aspects of orientifold planes that are important in order to elucidate their nature. 
For instance, in some cases  orientifold
planes turn out to have fractional charges with a positive or negative tension. For $Op$-planes with $p<6$
there are at least four different types of such planes \cite{wittenbaryons}, given by the non-trivial torsion values of suitable 
cohomology groups, 
i.e., by the possibility of turning on discrete fluxes of second rank anti-symmetric tensor fields (NS-NS or/and R-R).
These orientifold planes are labeled as $Op^+$, $Op^-$. \RPOp, \RNOp (see section 2 for details). 

Moreover, there is a brane setup
(see \cite{hananyo, sugimoto} and references therein) which establishes a mechanism to exchange one type of orientifold into a different
one. Basically the mechanism to describe the exchange $Op^- \leftrightarrow Op^+$ is to consider NS5-branes wrapping non-trivial 
(and suitable)
cycles in the transverse space $\IRP^{8-p}=\IR^{9-p}/\IZdos$.
This requires a quantization condition
for a field strength living on the worldvolume of the NS5-brane, as was proposed in \cite{sugimoto}. On the other hand, the exchange
\RNOp $\leftrightarrow$ \RPOp, is described by considering D$(p+2)$-branes wrapping non-trivial two-cycles of $\IRP^{8-p}$.

Although a classification of the above orientifold planes is provided by the cohomology groups of the transverse space\footnote{Actually, 
for an 
orientifold dimensionality less than two, cohomology classification of fields predicts at least 8 different types of orientifold planes. 
See \cite{bgs}} $\IRP^{8-p}$
\cite{bgs}, the fact that we have fractional charges for certain values of $p$, lead us to the conclusion that integral cohomology is not the
correct mathematical tool needed to classify orientifold planes. However, K-theory \cite{wittenk} turns out to be the correct 
one\footnote{Although K-theory 
is able to explain the relative charges among $Op$-planes, the question about the quantization condition for absolute fractional charges in
orientifolds still remains.}, as was shown in \cite{bgs} throughout a K-theory classification of RR fields.

However, in M-theory there are not gauge fields living on the worldvolume of solitonic objects. This means that K-theory is not
expected to be the relevant mathematical tool needed to classify objects there. In \cite{wittene8} a beautiful and detail computation in 
this sense was
done and the result was that cohomology is enough to classify objects in M-theory, at least for trivial geometrical backgrounds.

Hence,  one question that is immediately followed is if cohomology is enough to classify (without ambiguities) the M-theory lift of 
orientifold planes (this was studied in \cite{hananyo, gimon, sethi}, also see \cite{yo} for the M-theory lift of the orientifold six-plane)
and moreover, if the exchange between them can be realized (in M-theory) in a consistent way by using only the information cohomology
provides, i.e., {\it Is it possible to explain the relative charge of orientifold planes by an M-theory lift picture?} This is the problem 
we address in this note for the case of orientifold two-planes. We argue that the M-theory lift of 
orientifold two-planes is realized by considering $M5$-branes wrapping non-trivial three-cycles in the transverse projective space. 

We find that cohomology provides without ambiguity, the relative charge among the M-theory lift of orientifold planes by a two-fold
description: first, the relative charge is fixed by the cohomology groups (actually a relative cohomology group) and second, the flux
quantization condition that a M5-brane must satisfies (see \cite{fluxq}) naturally fixes the extra relative charge among these orientifold
planes. Hence we see that cohomology turns out to be the correct mathematical tool needed to classify and to describe the topological charged
objects in M-theory (or roughly speaking, up to now, cohomology seems to be enough). Moreover, if a cohomology classification in M-theory is
enough to classify the solitons in presence of `M-theory orientifold two-planes', the 
relative charge between 
\RNOtwo and $O2^-$ in type IIA theory, must be explained by the same picture (notice that in type II string theory, the mechanism to describe the
exchange of orientifolds $O2^+ \leftrightarrow O2^-$ is different that $\widetilde{O2^+}\leftrightarrow \widetilde{O2^-}$).

This is exactly what we obtain. We get a common picture in M-theory (a M5-brane wrapping a 3 cycle) that describes the exchange between pairs
of $OM2$-planes, and which is the `M-theory source' for  both orientifold-exchanging processes in type II string theory 
($O2^+ \leftrightarrow O2^-$ and $\widetilde{O2^+}\leftrightarrow \widetilde{O2^-}$) explaining in turn, their relative charge.

Also, we compute the flux quantization condition for the self-dual three-form field which lives on the worldvolume of the $M5$-brane and that,
at low energies, gives the flux quantization of the field strength for NS5-branes proposed in \cite{sugimoto}. 

It is important to point out, that
a formal calculation which proves that cohomology is the correct mathematical tool in M-theory  (in the presence of projective spaces),
as was done in \cite{wittene8}, is far for the aim of this note.

The outline is as follows: in section 2 we give a briefly review about orientifolds and the role that cohomology plays in their
classification. Also we describe the exchange of orientifolds in ten dimensions by considering branes wrapping homology cycles, and the 
quantization 
condition for the field strengths on the NS5-brane. In section 3, we describe the M-theory lift of the orientifold-exchange mechanism . 
We start by 
describing the M-theory lift of the orientifold two-plane. Afterwards, we review the action of the $M5$-brane and we identify
the relevant term needed to realize the exchange of the orientifolds. By using relative cohomology, we compute the quantization condition for the
self-dual field in the $M5$-brane worldvolume. This field (a three-form) give us the induced charge (in units of $M2$-brane) by integrating it
over a three-cycle. At the end of the section, we describe the M-theory picture which gives rise to the exchange of orientifold-two planes.

Finally, we give our conclusions in section four and in the appendix we describe a computation procedure in relative cohomology.

\section{The Exchange of Orientifold Planes in String Theory}
In this section we will briefly review some aspects of the orientifold exchange in string theory.
Basically the material is a review of \cite{hananyo, sugimoto, bgs, consistency}.

By starting from Type IIB theory it is possible to construct two different types of orientifold planes by gauging away the discrete symmetry
$\IZdos$ given by the orientifold projection. Hence, we can construct a negative RR charged nine-dimensional orientifold plane , denoted
by $O9^-$, or a positive charged one, denoted by $O9^+$. Besides the difference in the RR charges they carry, there are different fields
surviving the action of each plane. For instance, in the former case, the NS-NS two-from $B$ does not survive the projection, while in the
latter it does.

After T-duality, it is possible to construct orientifold planes with different dimensionalities. These planes, in Type IIA or IIB according 
to
their dimensionalities, are defined as
\beqa
Op^\pm\;=\;\IR^{p+1} \times \IR^{9-p}/\Omega \cdot {\cal J}\cdot {\cal I}_{9-p}\,,
\eeqa
where $\Omega$ is the usual parity projection on the string worldsheet (which reverses its orientation), ${\cal I}_{9-p}$ is the transversal
operator (reversing the sign on transversal coordinates to the orientifold plane) and ${\cal J}$ is equal to 1 for $p=0,1\,mod\,4$ and to
$(-1)^{F_L}$ for $p=2,3\, mod\, 4$. The RR charge for $Op^\pm$ is
$\pm 2^{p-5}$ respectively (in D$p$-branes units). The gauge groups related to the
worldvolume fields of $N$
D$p$-branes on top of them (and their images), are $SO(2N)$ for $Op^-$ and $USp(N)$ for $Op^+$.

These types of planes are valid for all values of $p$. However, for $p\leq 6$ there are other types of orientifold planes. Their existence is
suggested by the discrete torsion values given by the cohomology group $H^{6-p}(\IRP^{8-p};\IZ)=\IZdos$. It is also important to point 
out that there are two different types of fields (or forms): if the  field is
even under the projection, we say it is normal; if it is odd we call it a twisted field (form). Twisted cohomology
groups classify twisted forms.

The orientifold planes related to non-trivial discrete torsion values are denoted by
\RROp, and there are also two of them given by the positive or negative RR charge they carry. In the case of \RNOp the RR charge 
is equal to
$-2^{p-5}+\frac{1}{2}$, while in the case of \RPOp the RR charge is $+2^{p-5}$. The respective gauge groups associated to them are 
$SO(2N +1)$ and $USp(N)$.

The brane realization of these orientifolds has been very well studied \cite{hananyo, bgs}. For instance, the presence of \RROp-planes can be 
understood
by considering D$(p+2)$-branes intersecting $Op^{\pm}$-planes, or by wrapping D$(p+2)$-branes on homologically non-trivial and compact 
 2-cycles of
$\IRP^{8-p}$ \cite{yo2}. In the same token, by considering NS5-branes intersecting $Op^-$-planes or wrapping a non-trivial $(5-p)$-cycle of 
the transverse space $\IRP^{8-p}$, we obtain $Op^+$-planes and viceversa. These setups are reviewed in the next subsection.

Notice however that even though cohomology gives a quite correct classification of orientifold planes, there are two points where it fails: 
1) as we said above,
cohomology can not explain the origin of the relative charge among orientifold planes and 2) we require two cohomology groups in order to fix
completely the type of orientifold plane we are talking about. This is done 
by choosing the trivial or non-trivial values of the discrete valued cohomology groups. This means that it is necessary to 
consider the cohomology class of the NS-NS three-form and the cohomology class of the RR field strength $G_{6-p}\in\IZdos$ (besides the 
integer 
cohomology class $G_{8-p}\in H^{8-p}(\IRP^{8-p};\IZ)=\IZ$). Nevertheless, K-theory turns out to be the quite correct mathematical structure 
which solves the above two problems. It classifies the orientifold planes given a particular group for $Op^-$ and $Op^+$-planes which 
in turn takes into account the discrete values of RR fields which gives rise to the exotic orientifold planes \RROp. That is why the 
question about 
the utility of cohomology in M-theory is so important.

\subsection{NS5-branes and cohomology}
\begin{figure}
\begin{center}
\centering
\epsfysize=6cm
\leavevmode
\epsfbox{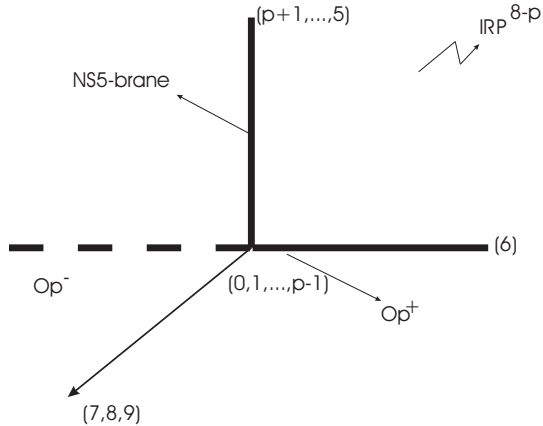}
\end{center}  
\caption[]{\small Brane realization of the exchange of $Op^-$ and $Op^+$.}
\label{NS}   
\end{figure}

In \cite{sugimoto} (also see \cite{consistency}), NS5-branes wrapping $\IRP^{5-p}$ were studied to explain the interchange of 
$Op^-$ and $Op^+$-planes. The basic
assumption is provided by a brane realization \cite{hananyo, bgs}. By taking a NS5-brane on coordinates 012345 and an $Op$-plane 
(positive  or negative) on 
$012\cdots p-1,6$, we
get the picture showed in figure \ref{NS}. 
The NS5-brane couples in a natural way to the field strength
$\tilde{B}_{(6)}$, which is the magnetic dual of the NS-NS two-form $B_{(2)}$. Due to the intersection of a NS brane and the orientifold plane,
we have a stuck half brane. Hence, the charge associated to this half-brane is given by the flux\footnote{$\IS^{n,m}$ stands for the
unitary sphere (of dimension $m+n-1$) of the space $\IR^{n,m}=\IR^m \times (\IR^n/\IZdos$).}
\beqa
\frac{1}{2\pi}\int_{\IS^{4,1}} H = \frac{1}{2} \:mod\;1\;,
\eeqa
where $H$ is the strength field $H=dB_{(2)}$. After using Stokes' theorem, we arrive at the quantization condition
\beqa
\frac{1}{2\pi}\oint_{\IRP^2} B\;=\;\frac{1}{2}\:mod\;1\;,
\eeqa
that is related to the discrete torsion cohomology $[\frac{H}{2\pi}]\in H^3(\IRP^{8-p};\ItZ)\;=\;\IZdos$. This holonomy 
contributes by a factor $g=e^{i\int_{\IRP^2}B}=-1$ in the $\IRP^2$ amplitude, and therefore exchanges the $Op^-$ and $Op^+$ planes.

The same picture can be obtained by wrapping NS5-branes on suitable homology cycles. To see that notice that $B_{(2)}$  is odd under the orientifold
projection, so it is a twisted form, classified by a twisted cohomology group. Hence, strings which couple to this two-form,
can only wrap twisted homological cycles. On the other hand, the NS5-brane can be wrapped on twisted or untwisted cycles according
to the orientifold dimensionality; this is because the dual six form $\tilde{B}_{(6)}$ will be a twisted form if the transverse
space to the orientifold is odd-dimensional, and untwisted or normal, if it is even-dimensional. This means that for odd (even) $p$, 
a NS5-brane can just
be wrapped on twisted (normal) cycles.

According to \cite{sugimoto}, a D$p$-brane RR charge is induced if the brane is wrapped on $(5-p)$-cycles. The homology group of the
corresponding cycles is given in table \ref{NSho}.

\begin{table}
\begin{center}
\caption{The homology classification of cycles where the NS5-branes can be wrapped on.}
\label{NSho}     

\begin{tabular}{||c|c|c||}\hline\hline
$Op$-plane&Homology group&Homology twisted group\\\hline\hline
$O5$&-&$H_0(\IRP^3;\ItZ)=\IZdos$\\\hline
$O4$&$H_1(\IRP^4;\IZ)=\IZdos$&-\\\hline
$O3$&-&$H_2(\IRP^5;\ItZ)=\IZdos$\\\hline
$O2$&$H_3(\IRP^6;\IZ)=\IZdos$&-\\\hline
$O1$&-&$H_4(\IRP^7;\ItZ)=\IZdos$\\\hline
$O0$&$H_5(\IRP^8;\IZ)=\IZdos$&-\\\hline\hline 
\end{tabular}
\end{center}
\end{table}

Following the notation in \cite{yo}, a ``NS$p$"-brane with topological charge
$\IZdos$ (given by the homology group) is obtained by wrapping a NS5-brane on $\IRP^{5-p}$. The RR charge is given by the term
\beqa
Q\;=\;\pm 2^{p-4} +\frac{1}{2} \:mod\;1\;,
\eeqa
where the second term in the rhs is provided by the presence of non-trivial torsion values of the cohomology group
$H^{5-p}(\IRP^{8-p};\IZ)=\IZdos$ and it is explicity given by the integral $\frac{1}{2\pi}\oint_{\IRP^{5-p}}C_{5-p}=\frac{1}{2}$.
Physically, we have an extra $\frac{1}{2} \:mod\;1$ term if $\IRP^{8-p}$ is the transverse space of an \RROp-plane.
The first term in the rhs comes from the flux quantization condition proposed in \cite{sugimoto} which reads,
\beqa
\frac{1}{2\pi}\oint_{\IRP^{5-p}} h_{5-p}\;\in\;\IZ + 2^{4-p}\;,
\label{conditionsNS}
\eeqa
where $h_{5-p}$ are the field strengths of the gauge fields on the NS5-brane, present in the NS5-brane action term
\beqa
\frac{1}{2\pi}\sum_p \int (h_{5-p} +C_{5-p})\wedge C_{p+1}\;,
\label{NSfields}
\eeqa
where $C$ are the RR fields and $p= 0,2,4$ for IIA theory and $p=-1,1,3,5$ for IIB theory.

Up to now we have seen that the discrete value of the cohomology group $H^3(\IRP^{8-p};\IZ)=\IZdos$ give us two types of orientifold planes,
$Op^-$ and $Op^+$. However  it does not seem obvious which variant would be identified with a trivial or non-trivial cohomology class. As was
shown in \cite{bgs}, K-theory turns out to be the correct mathematical structure which solves this problem. The K-theory group classifying
$Op^-$-planes is $KR^{p-10}(\IS^{9-p,0})$ and for $Op^+$ the K-theory group is $KH^{p-10}(\IS^{9-p,0})$ (see also \cite{yo2}). 
The use of K-theory as a classification tool for RR charges in string theory is natural, since the D-branes are naturally endowed with gauge
bundles. However, branes of M-theory are not and K-theory does not seem to arise in M-theory in the same natural way. The simplest proposal
is that charge under the M-theory forms is classified by cohomology. There are several possible avenues to try to extend our understanding of
the relation between K-theory in string theory and its M-theory lift. One of them is the question we address in this note.

If cohomology is the correct mathematical framework where charge in M-theory is classified, we must find an M-theory lift of the above
picture of orientifolds exchange and moreover, we must find that cohomology is enough to explain which variant of 'M-theory orientifolds' is
related to the trivial or non-trivial cohomology classes of the transverse space.
At the same time, we must reproduce the quantization condition (\ref{conditionsNS})  in terms of the M-theory fields.

\section{ The M-theory lift of the orientifold exchange}
In this section, we study the M-theory lift of the exchange of orientifold two-planes. However, before that, let us describe briefly some
important studies about the M-theory lifts of orientifold planes.

In particular, the lifting of the O4-plane has been studied in \cite{gimon} and also in \cite{consistency}. The lifting to M-theory of the
$O4$-plane is denoted as $OM5$ (see below). The lifting of $O0$, $O6$ and $O8$-planes are summarized in \cite{hananyo} (also see \cite{yo}
for the $O6$-plane) and connected by T-duality to the orientifold planes in Type IIB superstring theory. It is important to point out that 
the
quantization condition (\ref{NSfields}) has been successfully proved for the case $p=3$ (by S-duality) and for $p=4$ (see \cite{sugimoto}).

An M-theory picture of the exchange of orientifolds has been done for the case of the $O4$-plane \cite{consistency}. For
($p=0$) there is not a clear picture of the exchange mechanism (for $p=6,8$ there is not even a mechanism in ten dimensions). We shall focus 
on the
orientifold two-plane case, because it is the simplest case and because a flux quantization condition for a field in M-theory is 
directly followed. 

\subsection{Briefly review of M-theory lifts of orientifold planes}
The lifting of orientifold planes to M-theory has been studied in \cite{hananyo, gimon}. Particularly , we will concentrate on the study of 
the M-theory lift of the $O2$-plane. 

In M-theory, we have essentially the presence of membranes (two-branes) and their magnetic duals, five-branes. An `orientifold'
plane\footnote{Although there is not a worldsheet formulation, and hence the meaning of orientation reversal is lost, we will follow the
notation in \cite{hananyo} and we keep calling these planes `orientifold planes'.} in M-theory is defined as
\beqa
OMp\;=\;\IR^{p+1} \times \IR^{10-p}/\IZdos\;.
\eeqa
Here it is clear that the transverse space to the $OMp$-plane is the projective space $\IRP^{9-p}$. The action of the orientifold plane on
the fields is determined by invariance of the topological term in the action $\int C\wedge G\wedge G$, where $C$ is the three form which
couples to the two-brane, and $G=dC$. The three-form is transformed as $C \rightarrow (-1)^p C$ and it is required that
$p\;=\;1,2\;mod\;4$. Then the possible M-theory orientifold planes are: $OM1, OM2, OM5, OM6$ and $OM9$. By considering the $OM2$ and
$OM5$-planes, we obtain (upon compactification of the eleven coordinate) the well known $O2$- and $O4$-planes in Type IIA theory 
(see \cite{hananyo} 
for a description of the
rest of the O-planes).

As we are interested in the $O2$-plane, let us describe its M-theory lift in detail. The $OM2$-plane
is given by $\IR^{2,1}\times\IRP^7$ where the three form $C$ is invariant under the $\IZdos$ action, hence $C$ is 
a normal (no-twisted) 3-form. Being $\IRP^7$ an orientable space, the magnetic dual to $C$ is also a normal form (actually a 6-form which
couples to the five-brane). The charge of the $OM2$-plane is obtained by the term
\cite{sethi}: -$\int_{\IR^{2,1}\times\IRP^7}\,C\wedge I_8(R)$ and it is 
given by $Q\;=\;- 
\int_{\IRP^7}I_8(R)\;=\;-\chi/24$. It turns out that the Euler characteristic $\chi$ is 384, and after dividing by the 256 fixed points, 
we obtain 
that $Q\;=\;-\frac{1}{16}$. On the other hand, the only field strength in the bulk is  $G$, so we are interested in the cohomology 
group $[\frac{G}{2\pi}]\in
H^4(\IRP^7;\IZ)=\IZdos$. However, this non-trivial discrete torsion value, can not be expressing the fact that we have half fluxes for $G$,
i.e., that $\int_{\IRP^4}\frac{G}{2\pi}\;=\;\frac{1}{2}$, since this contradicts the flux quantization condition for $G$ studied in
\cite{fluxq}, where $2\int\frac{G}{2\pi}\;=\;\int\omega_4 \;mod\;2$, and $\omega_4=0$ for $\IRP^7$. 

The term associated to the above discrete torsion, as was computed in \cite{sethi}, is 
\beqa
-\frac{1}{2}\int_{\IRP^7} \left(\frac{1}{(2\pi)^2}\right)G\wedge
C=\frac{1}{4}\;,
\eeqa
 and it is which gives an extra charge to the OM2-plane. Hence, we have two versions of $OM2$-planes denoted by
$OM2^-$ (related to the trivial value of the above cohomology group), and $OM2^+$, (related to the non-trivial one). The charge of $OM2^-$ is
$Q_-= -\frac{1}{16}$ and the one for $OM2^+$ is $Q_+=-\frac{1}{16} +\frac{1}{4}=\frac{3}{16}$.

In type IIA string theory we have four versions of O2-planes. The M-theory lifts of
these planes, is as follows:
\begin{itemize}
\item
An $O2^-$-plane ($Q=-\frac{1}{8}$) is lifted to a pair of $OM2^-$-planes, on the two fixed points on $\S^1$ (the compact eleven direction).
\item
An $O2^+$-plane ($Q=+\frac{1}{8}$) is lifted to a pair  ($OM2^-$,$OM2^+$).
\item
An \RPOtwo-plane ($Q=+\frac{1}{8}$) has a M-theory lift given by the pair ($OM2^+$,$OM2^-$) (this setup differ from the last one in the
location of the two $OM2$-planes).
\item
An \RNOtwo-plane ($Q=-\frac{1}{8}+\frac{1}{2}=\frac{3}{8}$) is obtained from M-theory by a pair of $OM2^+$-planes.
\end{itemize}

As we have seen in the previous section, in Type II superstring theory we can exchange the two types of orientifold $Op^- \leftrightarrow
Op^+$ or \RNOp- $\leftrightarrow$ \RPOp planes by a NS5-brane intersecting the orientifold plane or by wrapping it on a non-trivial
compact homological $(5-p)$-cycle. The question that immediately follows is: {\it can we exchange the
two $OM2$-planes by considering a M5-brane wrapping a suitable homological cycle in $\IRP^7$?} If cohomology is the proper mathematical 
tool needed for M-theory,
as was shown in \cite{wittene8}, the answer must be affirmative. We study this setup in the next section. 

\subsection{M5-brane action}
\begin{figure}
\begin{center}
\centering
\epsfysize=6cm
\leavevmode
\epsfbox{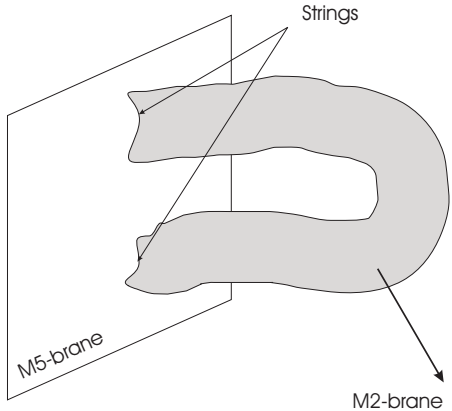}
\end{center}  
\caption[]{\small A solitonic configuration in M-theory: an $M2$-brane can be attached to an $M5$-brane.}
\label{m2m5}   
\end{figure}

The action terms for the M5-brane have been well studied in the last years (see for instance \cite{branes, M5schwarz,elevenm5, 
townsend, otroschwarz, schwarz3,
ganomaly, ganomaly2, BPSm5, m5action} and \cite{sorokin, sorokin2} for the original papers where the action for the M5-brane was constructed
in a generic bosonic background and the supersymmetric case, respectively). The bosonic content involves the metric and the three-form,
and it is given by the 11D supergravity action,
\beqa
S_1 \sim \int_{M_{11}} G\wedge *G -\frac{1}{6}C\wedge G\wedge G\;.
\eeqa
There is also a kinetic term given by
\beqa
S_2 \sim \int_{M_{11}}R +\tau_{M_5}\int_{W_6} (det G)^{1/2} +\tau_{M_2}\int_{W_3} (det\,G)^{1/2}\;,
\eeqa
where $W_6$ and $W_3$ are the worldvolumes of M5 and M2 branes respectively. Besides these ones, the
consistency of the theory requires the presence of the so called Wess-Zumino terms,
\beqa
S_3 \sim \int_{W_7} C\wedge G\;+\;\int_{W_8}G\wedge G
\eeqa
(notice that the first term in the WZ term is the responsible for giving an extra charge to the M2-brane in the presence of an $OM2$-plane). 
On the other hand, in M-theory we also have some solitonic configurations. These are for instance, the ending of a M2-brane on a M5-brane, which
is the M-theory realization of the fact that in Type IIA superstring theory, D2-branes end on NS5-branes. A M2-brane ending on a M5-brane
(see figure \ref{m2m5}) establishes a one-dimensional submanifold of M5. This string must couples naturally to a two-form, but there are not two
forms in the whole eleven dimensional spacetime. However, as was studied in \cite{M5schwarz, townsend, otroschwarz, schwarz3}, the theory 
requires 
the existence of a self-dual field strength $T^\dag_{3}=dA^\dag_{2}$  (a three-form).

This three form lives on the worldvolume of the M5-brane and it is represented in the M5-brane action by the Chern-Simons term,
\beqa
S_4 \sim \int_{W_6} T^\dag_3 \wedge *T^\dag_3 + T^\dag_3\wedge C_3\;.
\eeqa
Finally we get the topological couplings of the fields $C_3$, $\tilde{C}_6$ and $A^\dag_2$ to $W_6$, $W_3$ and $W_2$, where $W_2$ stands for
the worldsheet spread out by the string in M5,
\beqa
S_5 \sim \int_{W_6}\tilde{C}_{(6)} + \int_{W_3}C_3 + \int_{W_2} A^\dag_2\;.
\eeqa
Notice that the last factor in the Chern-Simons term is counting twice for the self-duality of the fields \cite{largegauge, m5action}.

\subsection{The cohomology of $T^\dag_{(3)}$}

According to \cite{largegauge}, the classification of $T^\dag_{(3)}$ is given by the relative cohomology group $H^4(X,W_6;\IZ)$,
where $X$ is the eleven dimensional spacetime. In general, relative cohomology (see \cite{largegauge} for a very good exposition about 
relative cohomology in
the case of M5-brane and \cite{rcohomology} for an application to string theory\footnote{Also see \cite{hatcher} for a mathematical
exposition.}) classifies
$k$-forms $\Lambda_k$ satisfying the following assertions:
\beqa
d\Lambda_k=0\\\nonumber
i*\Lambda_k =0
\eeqa
where $i: W \rightarrow X$ is the inclusion of the subspace $W$ to the space $X$, and the relative cohomology group is given by
$H^k(X,W;\IZ)$. In the case we have a non-trivial $\lambda_{k-1}$-form living in the subspace $W$, the conditions read
\beqa
\begin{array}{ccc}
d\Lambda_k&=&0\\
i*\Lambda_k&=&d\lambda_{k-1}
\end{array}
\eeqa
where $d\lambda_{k-1}=0$. The notation is $[(\Lambda_k, \lambda_{k-1})]\in H^k(X,W;\IZ)$ and moreover, there is a relation to the de Rham
cohomology given by
\beqa
\int_{\Sigma_k, \sigma_{k-1}}(\Lambda_k, \lambda_{k-1}) \;=\;  \int_{\Sigma_k}\Lambda_k \;-\; \int_{\sigma_{k-1}} \lambda_{k-1}\;,
\eeqa
where $\Sigma_k$ is a homological $k$-cycle and $\partial\Sigma_k=\sigma_{k-1}$. 

For the $M5$-brane in M-theory, we have  a 4-form $G$ and a 3-form $T^\dag_3$ which is living on the worldvolume of the
five-brane. Hence, the suitable cohomology group for these forms must be the relative cohomology group $H^4(X,W_6;\IZ)$. However in the
problem we address in this note, the presence of an orientifold plane plays an important role. Let us describe the situation in detail.

\begin{figure}
\begin{center}
\centering
\epsfysize=6cm
\leavevmode
\epsfbox{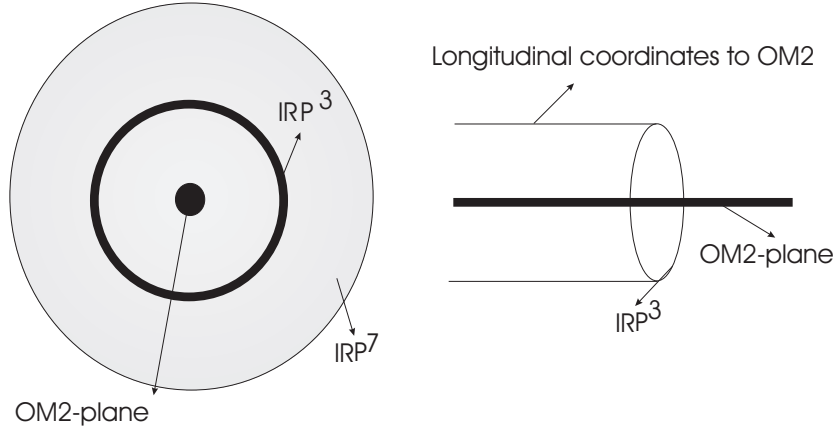}
\end{center}  
\caption[]{\small Schematic picture of the wrapping of $M5$ on $\IRP^3$.}
\label{OM2}   
\end{figure}

According to section 2, the M5-brane is intersecting the orientifold OM2-plane at coordinates 01. This means that such coordinates are fixed
on the orientifold plane and that we have 4 more coordinates to change the configuration by wrapping the M5-brane on a homological non-trivial
three-cycle of $\IRP^7$. As the M5-brane is embedded in an eleven dimensional space, we have the freedom to wrap the coordinates 
2345 on $\S^3$ (in the covering space) $\times \IR$ (with one of them being longitudinal to $OM2$). This picture show us 
that the 
M5-brane worldvolume can be described by
$\IR^{1,1}\times \IR\times \IRP^3$(see figure \ref{OM2}). In conclusion, the relative cohomology group which classifies the forms 
$(G_4,T^\dag_3)$ is 
$H^4(\IRP^7,\IRP^3;\IZ)$.
The computation of this group is shown in the appendix\footnote{I thank to A. Pedroza for a detailed explanation of this.} and the result 
is that 
$H^4(\IRP^7,\IRP^3;\IZ) =\IZ$.\\

\subsection{Flux quantization condition for $T^\dag_3$}
Now we are ready to compute the flux of $T^\dag_3$ around $\IRP^3$. The physical meaning of the relative cohomology value 
$H^4(\IRP^7,\IRP^3;\IZ)=\IZ$, is as follows. According to the appendix, a half integer flux is induced on $T^\dag_3$ by the non-trivial 
class in $H^4(\IRP^7;\IZ)=\IZdos$, i.e., by the presence of the field strength $\left[\frac{G}{2\pi}\right]\in H^4(\IRP^7;\IZ)$. 
Therefore, the flux of $T^\dag_3$ is shifted to
\beqa
\begin{array}{cccc}
\oint_{\IRP^3}\left(\frac{1}{2\pi}\right)T^\dag_3&\in&\IZ\oplus\IZ/2&\\
&=&n+\frac{2k+1}{2}\,&=\,l+\frac{1}{2}
\end{array}
\label{anterior}
\eeqa
where $k,l,n \in \IZ$.  As we have seen, however, the three-form $T^\dag_3$ is self-dual. This means that up to now we are 
considering electric and magnetic parts for it. In order to obtain the desired flux quantization condition, for self-dual
fields, we must divide by two\footnote{This is also supported by the formula (\ref{conditionsNS}) for $p=2$ in Type IIA 
string theory, where the field $h_3$ is not self-dual and its magnetic part is not considered.} \cite{m5action, largegauge}. Hence, 
the correct quantization condition for the self-dual field $T^\dag_{(3)}$ reads\footnote{$\frac{1}{2}(l+\frac{1}{2})=\frac{l}{2}+\frac{l}{4}$.
If $l=2q$ with $q\in \IZ$, then equation (\ref{anterior}) is equal to $q +\frac{1}{4}$. If $l=2q+1$, we have that equation (\ref{anterior}) 
is equal to $(q+1)-\frac{1}{4}$. Since we are interested only in fractional charges, we do not consider the interpretation of this extra unit
$M2$ charge.}
\beqa
\oint_{\IRP^3}\left(\frac{1}{2}\right)\frac{T^\dag_3}{2\pi}\,\in\,\IZ\pm\frac{1}{4}\;.
\label{penul}
\eeqa
Notice that at this point, it seems that we have the same problem as with orientifolds at the string theory level since $T_3^\dag$ is
classified by $H^4(\IRP^7;\IZ)$: it does not seem clearly
which variant of charge (+1/4 or -1/4) would correspond to which type of orientifold plane. We could have an $M5$-brane wrapping $\IRP^3$
around an $OM2^-$ or an $OM2^+$, so which charge is given to the $OM2$-plane, $-1/4$ or $+1/4$?
In other words, the presence of the discrete torsion
$H^4(\IRP^7;\IZ)=\IZdos$ give us the possible charges in units of $M2$-branes that a $M5$-brane carries when it wraps a 
three-cycle. Naively
we can argue that cohomology does not give us such an information, but it does as we see in the next subsection.

Let us give a second argument that supports the above assertion. As it is shown in the appendix, the relative cohomology group is 
integer valued. This means that
\beqa
\frac{1}{2\pi}\int_{({\cal W}, \IRP^3)}(G, T^\dag_{(3)})\;=\;\frac{1}{2\pi}\int_{\cal
W}G_4\;-\;\frac{1}{2\pi}\int_{\IRP^3}T^\dag_{(3)}\;=\;n \:\:\:n\in\IZ\;,
\label{relative}
\eeqa
where ${\cal W}$ is a 4-cycle in $\IRP^7$ with boundary $\partial{\cal W}=\IRP^3$. Due to the fact that 
$\left[\frac{G}{2\pi}\right]\in H^4(\IRP^7;\IZ)=\IZdos$ 
we actually have two cohomological classes for the field $G$\footnote{The field $G$ satisfies in turn a flux quantization condition 
\cite{sethi, consistency, fluxq} (also see \cite{fivebranesw}). In particular, the flux of $G$ over a 4-cycle is related to the 
Stiefel-Weyl four-class
$\omega_4$, by
\beqa
\frac{1}{2\pi}\int_{\cal W}G\;=\;\int\omega_4\;mod\,2\;,
\eeqa
but for the $\IRP^7$ case, $\omega_4$ is zero \cite{fluxq}, and then it is not possible to have half fluxes for $G$. Therefore, 
$\left[\frac{G}{2\pi}\right]\in\IZdos$ implies that $\int_{\cal W}\frac{G}{2\pi}\,=\,n\,mod\,2$, with $n\in\IZ$.}.
The trivial one, $\left[\frac{G}{2\pi}\right]=[0]$, implies that the field $T^\dag_3$ in equation (\ref{relative}) reads
\beqa
\frac{1}{2\pi}\int_{\IRP^3}T^\dag_{(3)}\;=\;n\:,
\eeqa
while the non-trivial class, $\left[\frac{G}{2\pi}\right]=[1]$ implies that
\beqa
\frac{1}{2\pi}\int_{\IRP^3}T^\dag_{(3)}\;=\;n + \frac{1}{2}\:.
\eeqa
This can be straightforward read it off from the spectral sequence given in the appendix. Afterwards, by considering the self-duality of 
the strength field $T^\dag_3$, we get the flux quantization condition (\ref{penul}).

\subsection{The M-theory lifting picture}
In this section we finally describe the M-theory lift of the exchange  $O2^+ \leftrightarrow O2^-$. The 
setup in string theory where a NS5-brane intersects an $O2$-plane, is lifted to M-theory as a pair of $OM2$-planes 
with one of them intersecting the M5-brane. 

Take for instance an $OM2^-$-plane on coordinates 016 and a M5-brane on coordinates 012345 in the eleven dimensional spacetime. The
transverse space to the orientifold plane is $\IRP^7$ and we can wrap a M5-brane on non-trivial homological cycles of this space. We are
interested in wrapping the M5-brane on a three cycle, according to the picture in ten dimensions. The three-cycles are classified by the
homology group $H_3(\IRP^7;\IZ)=\IZdos$ (recall that M5 must be wrapped on normal cycles). The non-triviality of this group enable us to
wrap the five-brane on such cycles.

\begin{figure}
\begin{center}
\centering
\epsfysize=6cm
\leavevmode
\epsfbox{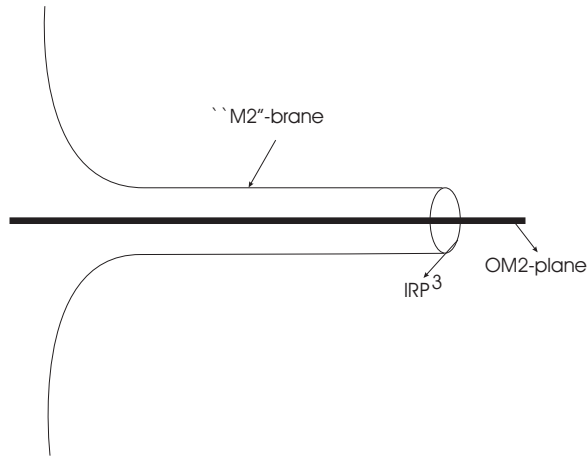}
\end{center}  
\caption[]{\small  A $M5$-brane wrapping $\IRP^3$ transverse to an $OM2$-plane.}
\label{wrapping}   
\end{figure}

After wrapping the M5-brane on $\IRP^3$ (see figure \ref{wrapping}), naively, a ``$M2$''-brane with topological charge
$\IZdos$ is obtained. This means that if we wrap another M5-brane on $\IRP^3$, the total effect is null. In this context, an $OM2^-$-plane changes to an
$OM2^+$-plane when a M5-brane is wrapped on $\IRP^3$ and returns back to an $OM2^-$-plane when another M5-brane is wrapped on the
three-cycle.

However, we must explain the origin of the $M2$-brane charge which is provided when a $M5$-brane is wrapped on such a cycle. The
relevant factor to consider is the Chern-Simons term $\int_{W_6} C\wedge \left(\frac{1}{2}\right)T^\dag_3$. After wrapping the M5-brane on a 
three cycle, this term reads,
\beqa
\int_{\IR^{2,1}\times\IRP^3}C\wedge \left(\frac{1}{2}\right)T^\dag_3\;=\;\int_{\IR^{2,1}}C \cdot \int_{\IRP^3}\left(\frac{1}{2}\right)T^\dag_3\;.
\eeqa
Notice that we have only considered  the electric flux given by the self-dual field $T^\dag_3$. Hence the M2-brane charge $Q_{M2}$ is given by 
\beqa
Q_{M2}\,=\,\int_{\IRP^3}\left(\frac{1}{2}\right)T^\dag_{(3)}\;=\;\pm\frac{1}{4}\;,
\eeqa
according to the last subsection\footnote{Where we are taking the minimal quantization value.}.

The conclusion is that by wrapping a M5-brane on $\IRP^3$, a charge equal to $\pm 1/4$ (in units of M2-brane charge) is induced. Also, 
the flux quantization condition for the field $h_3$ in equation
(\ref{NSfields}) is explained (for the
case $p=2$). The M-theory lift of $h_3$, is the self-dual field strength $T^\dag_3$ living in the worldvolume of the M5-brane; this in
agreement with the pictures we already had: the existence of the $h_3$ field living in the worldvolume of the NS5-brane in 
ten dimensions follows from the fact that
D2-branes end on NS five-branes. In the M-theory lift, the $h_3$ field corresponds to the self-dual field
$T^\dag_3$ which  is a consequence of the fact that M2-branes can be attached to M5-branes. 

Now, let us describe the exchange of $OM2$-planes in M-theory and also the way that cohomology establishes a difference between the charges
that they acquire (i.e., in which case, a positive or negative 1/4 charge, must be taken into account).
\begin{itemize}
\item
$O2^- \leftrightarrow O2^+$ lifts to M-theory to the exchange of $(OM2^-, OM2^-) \leftrightarrow (OM2^+, OM2^-)$. This is explained by
wrapping a M5-brane on $\IRP^3$ on one of the two $OM2^-$-planes. The charge is now $-1/16 + 1/4=3/16$, i.e., the $OM2^-$-plane acquires an
extra $+1/4$ charge.
\item
\RPOtwo $\leftrightarrow$ \RNOtwo, is lifted to the exchange $(OM2^+, OM2^-) \leftrightarrow (OM2^+, OM2^+)$. This is explained by wrapping
a $M5$-brane on the three-cycle on the negative $OM2^-$-plane. The orientifold plane acquires a positive $+1/4$ charge. The opposite
situation is an $OM2^+$-plane getting an extra negative charge of $-1/4$.
\item
Consider now the possibility of wrapping a $M5$-brane on a three-cycle on $OM2^-$ such that, according to the results of the last subsection, 
acquires a
$-1/4$ charge. The total charge must be\footnote{Of course, there are infinite ways to decompose $-5/16$, but we are interested in
decompose it in terms of the known $OM2$ charges.} $Q=-1/16-1/4=-5/16=3/16-1/2$. This is interpreted as having an $OM2^+$ and a stuck half $M2$-brane. But a
single $M2$-brane intersecting the $OM2$-plane is not allowed by the flux quantization condition of $G$ (see \cite{hananyo, consistency}). Hence this situation
is not possible (we could avoid the presence of a half $M2$-brane by wrapping other $M5$-brane, but this give us a null net charge). The same
situation figures out when we consider an $OM2^+$-plane acquiring a positive $+1/4$ charge. 
\end{itemize}
Notice that it is possible to explain the exchange of an $OM2^+$ to an $OM2^-$ by  taking the former one as the result of exchanging an
$OM2^-$-plane into it and recalling that a zero total charge is obtained by wrapping twice a M5-brane on $\IRP^3$. More important,
notice also that cohomology turns out to be the correct mathematical tool needed to describe the above situations, since it gives 
sufficient information (relative charges and flux quantization condition for $G$) which allows us to elucidate
which orientifold plane is related to a positive or negative extra charge.

\subsection{The exchange of $O2^\pm \leftrightarrow$ \RROtwo}
The next question is to construct the M-theory lift of the exchange of  orientifold planes related to the non-trivial discrete value of
the cohomology group $H^4(\IRP^6;\IZ)=\IZdos$, i.e., the orientifold planes denotes as \RROtwo. It is well known \cite{hananyo, sugimoto,
bgs}, that the exchange of
$O2^-$ to \RNOtwo  ($O2^+$ $\leftrightarrow$ \RPOtwo) is given in type IIA string theory by the intersection of a D4-brane and an orientifold
two-plane ,
or in other words, by wrapping the D4-brane on a non-trivial compact homological two-cycle classified by the homology group
$H_2(\IRP^6;\ItZ)=\IZdos$ Ponicar\'e dual of the above cohomology group.

\begin{figure}
\begin{center}
\centering
\epsfysize=6cm
\leavevmode
\epsfbox{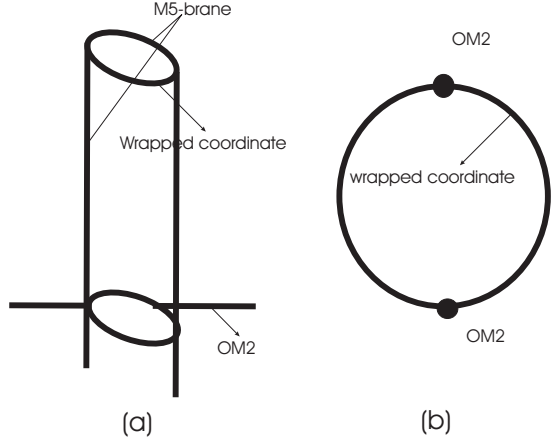}
\end{center}  
\caption[]{\small a) The $M5$-brane must be wrapped on the same compact direction where the eleven dimension is compactified. b) 
In the same wrapped coordinate we actually have two $OM2$-planes.}
\label{2OM2}   
\end{figure}

The simplest M-theory lift of this setup is given by intersecting at least one of the two $OM2$-planes and the $M5$-brane. The $M5$-brane
reduces upon a circle compactification to a D4-brane in type IIA string theory, hence such a compactification must be taken along one of the
$M5$-brane coordinates. On the other hand, as we saw, there are indeed two $OM2$-planes fixed in two different points in such a coordinate.
The resulting picture is that we actually have a $M5$-brane intersecting both $OM2$-planes. See figure \ref{2OM2}.

\begin{figure}
\begin{center}
\centering
\epsfysize=6cm
\leavevmode
\epsfbox{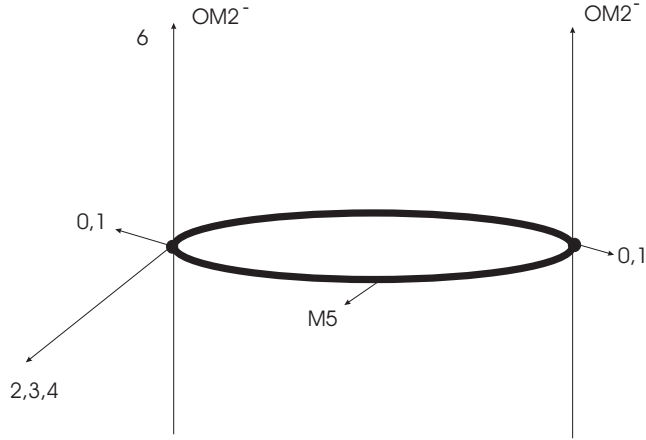}
\end{center}  
\caption[]{\small The M-theory lift of the exchange $O2^\pm \leftrightarrow \widetilde{O2^\pm}$.}
\label{last}   
\end{figure}

But this is indeed what our picture about the exchange of $O2^\pm \leftrightarrow O2^\mp$ give us. In
such a picture we have that by wrapping a M5-brane on a three cycle, the charge of the $OM2$-plane is changed. If we wrap two $M5$-branes on
three-cycles around each of the two $OM2$-planes (before compactification), there will be a change in the charge carried by both of them. 
Hence, by wrapping a $M5$-brane on each
orientifold plane, we get
\begin{itemize}
\item
The pair $(OM2^-, OM2^-)$ (the M-theory lift of an $O2^-$-plane) transforms into  the pair $(OM2^+, OM2^+)$, which is the M-theory lift of 
an \RNOtwo-plane, and viceversa.
\item
$(OM2^-, OM2^+)$, the lifting to M-theory of an $O2^+$-plane, transforms into the pair $(OM2^+, OM2^-)$, which is the lifting of an
\RPOtwo-plane, and viceversa.
\end{itemize}
Hence, the M-theory lift of the exchange $O2^\pm \leftrightarrow$ \RROtwo, is given  by wrapping the same $M5$-brane on both
orientifolds once such a coordinate is compact. See figure \ref{last}.

\section{Conclusions}
In this note we study the M-theory lift of the exchange of orientifold two-planes. In type IIA string theory, the exchange mechanism 
for $O2^-$ going to $O2^+$,  involves the
wrapping of a NS5-brane on $\IRP^3$, where the three-cycle is classified by the homology group $H_3(\IRP^6;\IZ)=\IZdos$. In order to 
explain the
extra RR charge acquired by the orientifold plane, it was proposed in \cite{sugimoto} that the field $h_3$ (present in the action  of the
NS5-brane) must satisfies a flux quantization condition. In particular that $\oint_{\IRP^3}\frac{h_3}{2\pi}\in \IZ+1/4$. The origin for this
condition was not well understood. Although cohomology properly classifies the orientifold planes in string theory, it does not explain the 
relative discrete charge among, e.g., \RNOtwo and $O2^-$ planes, and moreover, it is not possible to know which orientifold variant, $O2^+$ 
or $O2^-$ is related to which cohomology class of $H^3(\IRP^{6};\IZ)=\IZdos$. These are some of the reasons to consider K-theory instead 
of cohomology. Nevertheless, K-theory does not seem to play a role in the classification of charges in M-theory. The question we address in 
this note is to check if cohomology is sufficient to explain the M-theory lifts of the above features in Type IIA string theory.

The M-theory lift of the exchange of orientifolds that we propose is the wrapping of the $M5$-brane on $\IRP^3$ on the transverse
space to an $OM2$-plane. The M-theory lift of an $O2$-plane is a pair of $OM2$-planes. The extra $M2$-brane charge that the
orientifold plane in M-theory acquires by the above wrapping is given by the term 
$\frac{1}{2\pi}\oint_{\IRP^3}\left(\frac{1}{2}\right)T^\dag_3\in \IZ\pm 1/4$, where
$T^\dag_3$ is the self-dual field living on the worldvolume of the $M5$-brane, and corresponds to the field strength of the two-form field
$A_2$ which in turn couples the one-dimensional region where a $M2$-brane ends on a $M5$-brane. We argue that this is the origin of the
quantization condition proposed in \cite{sugimoto}. This was computed by using the relative cohomology group $H^4(X,W_6,\IZ)=\IZ$. 

This picture, however, lead us to the same problem we had in the cohomology classification of orientifolds in Type IIA string theory. 
We actually have two possible charges, $\pm 1/4$, to assign to each M-theory orientifold plane $OM2$. The solution to this problem, 
arises by considering the flux quantization condition for $G$. Such a condition, in the presence of an M-theory orientifold plane,  
prohibits us to give a negative (positive) extra charge to a negative (positive) $OM2$ plane. In this way, cohomology turns out to be 
sufficient (at least up to this case) to classify (without ambiguity) the charges of orientifold planes and at the time, it is possible 
to give an explanation for the relative charges among them.

In the procedure, we also report that the M-theory lift of the exchange of orientifolds $\widetilde{O2^\pm} \leftrightarrow O2^\pm$ has the 
same picture as 
the above case, i.e., the wrapping of a M5-brane on $\IRP^3$. This is important, because we do not
require other mathematical tools to distinguish between $O2^\pm$-planes and $\widetilde{O2^\pm}$-planes as was done for orientifolds in 
string theories, 
where the later were constructed by wrapping a $D4$-brane on $\IRP^2$ (and classified by $H_2(\IRP^6;\IZ)-\IZdos$).

It would be interesting to study the exchange mechanism of orientifolds for the case of $O0$ and in general the wrapping of solitonic
objects in M-theory on homological cycles related to more complicated geometries. We hope that this note could be useful for future 
research on the mathematical structure of
orientifolds and M-theory.


\begin{center}
{\bf Acknowledgments}
\end{center}

It is my pleasure to thank Amihay Hanany for his hospitality and for very useful suggestions and the Center of Theoretical Physics of MIT for kind
hospitality. Also I would like to thank N. Constable and A.
Pedroza for very fruitful discussions and especially to H. Garc\'{\i}a-Compe\'an and A. Uranga for great support and useful conversations. 
This work is
supported in part by a CONACyT (M\'exico) fellowship with number 020136 and by the U.S. Department of Energy (D.O.E.) under cooperative
research agreement \#DF-FC02-94ER40818. 

\hspace{0.5cm}


\appendix
\section{ Computation of $H^4(\IRP^7,\IRP^3;\IZ)$}
In this appendix  we show the procedure to compute the cohomology group $H^4(\IRP^7,\IRP^3;\IZ)$. 

It is said that $(\Omega_k, \omega_{k-1}) \in H^k(X, W;\IZ)$ if, 
\beqa 
\begin{array}{ccc} 
d\Omega_k&=&0\\ i*\Omega_k&=&d\omega_{k-1}
\end{array} 
\eeqa 
where $i:W\rightarrow X$ is the inclusion. In the dual picture, i.e., in the relative homology,  it is said that a
$k$-cycle $\Sigma_k$ is non-trivial if it is not the boundary of some submanifold of $X$, except in $W$. This means that there is a
$(k-1)$-cycle on $W$ such that 
\beqa 
\partial\Sigma_k = \sigma_{k-1}\in W\,. 
\eeqa 
Now consider the long exact sequence of cohomology
groups: 
\beqa 
\cdots \rightarrow H^k(X,W;\IZ) \rightarrow H^k(X;\IZ)\rightarrow H^k(W;\IZ) \rightarrow H^{k+1}(X,W;\IZ)\rightarrow \cdots\;.
\eeqa 
The knowledge of some of the groups involved in the above sequence can allowed us to compute other one. We show this with the example
that is of interest for us: the calculation of the relative cohomology group $H^4(\IRP^7,\IRP^3;\IZ)$. The long exact sequence turns to be
{\small 
\beqa 
\cdots\rightarrow H^3(\IRP^7;\IZ)\rightarrow H^3(\IRP^3;\IZ)\rightarrow H^4(\IRP^7,\IRP^3;\IZ)\rightarrow
H^4(\IRP^7;\IZ)\rightarrow H^4(\IRP^3;\IZ)\rightarrow\cdots 
\eeqa 
} 
The cohomology groups of the projective spaces are known (see \cite{hananyo}), and the sequence becomes, 
\beqa 
\diagram{
\cdots&\harr{}{}&0&\harr{}{}&\IZ&\harr{j^*}{}&H^4(\IRP^7,\IRP^3;\IZ)&\harr{p^*}{}&\IZdos&\harr{}{}&0\harr{}{}&\cdots\;, 
} 
\eeqa 
where the mappings are the coboundary map $j^*\omega_{k-1}=(0,\omega_{k-1})$, the projection 
$p^*(\Omega_k, \omega_{k-1})=\Omega_k$ and the
pullback of the inclusion $i$.
Hence, we have a
short exact sequence. A short exact sequence (see \cite{hatcher}) is given by 
\beqa 
\diagram{
0&\harr{}{}&A&\harr{j^*}{}&B&\harr{k^*}{}&C&\harr{}{}&0 } 
\eeqa 
where $Im\,j^*\;=\;Ker\,k^*$. This implies that $C=B/A$. Hence in our case, we have
that 
\beqa 
\IZdos\,=\,H^4(\IRP^7,\IRP^3;\IZ)/\IZ \;.
\eeqa 
The value of the relative cohomology group, can be easily read it from the following long exact sequence of chains:
{\small
\beqa
\diagram{
0&\harr{}{}&C_7=\IZ&\harr{0}{}&C_6=0&\harr{\times 2}{}&C_5=\IZdos&\harr{0}{}&C_4=0&\harr{\times 2}{}&\\\nonumber
C_3=\IZdos&\harr{0}{}
&C_2=0&\harr{\times 2}{}&C_1=\IZ&\harr{0}{}&C_0=\IZ&\harr{}{}&0}
\eeqa
}
and the result is that $H^4(\IRP^3,\IRP^7;\IZ)=\IZ$. This means that $j^*= \times 2$ and physically this implies that a half-integer flux is
induced on $H^3(\IRP^3;\IZ)$ (see \cite{bgs}). To see this (in an informal way) notice that an integer flux in $H^3(\IRP^7,\IZ)=\IZ$ is mapped to twice an integer element in the relative cohomology group $H^4(\IRP^7,\IRP^3;\IZ)$ which in turns maps to a class in $\IZdos$ ($=H^4(\IRP^7;\IZ)$). Hence, we see that even fluxes in the relative cohomology flux are mapped to the trivial class in $\IZdos$, while odd fluxes are mapped to the non-trivial class in $\IZdos$. In order to obtain odd and even fluxes in the relative cohomology group, we need an element flux in $H^3(\IRP^7;\IZ)$ such that under multiplication by two, give us and odd flux. The conclusion is that we require a half-integer shift in the fluxes classified by $H^3(\IRP^7;\IZ)$.

\end{document}